\documentclass[preprint]{elsarticle}
\usepackage{graphicx}
\usepackage{amssymb}
\begin{document}

\title{Vacuum Annealed Cu contacts for graphene electronics}
\author{C.~E.~Malec,~B.~Elkus,~D.~Davidovi\'c}
\address{Georgia Tech Physics Dept}

\begin{abstract}
We present transfer-length-method measurements of the contact resistance between Cu and graphene, and a method to significantly reduce the contact resistance by vacuum annealing. Even in samples with heavily contaminated contacts, the contacts display very low contact resistance post annealing. Due to the common use of Cu, and itÕs low chemical reactivity with graphene,  thermal annealing will be important for future graphene devices requiring non-perturbing contacts with low contact resistance.
\end{abstract}
\date{\today}
\maketitle

\section{Introduction and Fabrication}
Many electronic devices rely on Cu interconnects to other circuit components.  Future applications of graphene such as graphene transistors would require an understanding of properties of contacts made to graphene.~\cite{nagashio}    Due to its ubiquity in current electronic architecture and weak chemical interaction with graphene,~\cite{giovannetti,barraza} the ability to create Cu leads to graphene with contact resistances comparable to contacts of Pd or similar materials~\cite{xia,huard,lee,blake,khomyakov} may prove critical to future graphene applications.  Here we measure the contact resistance of Cu leads to graphene using the transfer length method (TLM).  Two devices (A \& B) are presented in this paper, however; we have performed less extensive tests on other samples.

Mechanical exfoliation is used to deposit graphene onto the surface of an oxidized silicon wafer.~\cite{novoselov0}  Single graphene layers are identified using an optical microscope, and  authenticated via Raman spectroscopy.~\cite{ferrari}  Our devices were created by two e-beam lithography processes.  The first was used to define a Poly(methyl methacrylate) (PMMA) mask to protect the graphene from an O$_2$ plasma to create well defined rectangular strips. All remaining PMMA is removed using Acetone. In the second process, we use a methacrylic acid/PMMA bilayer of resist and define exposed regions for metal contacts.  The leads of sample B were slightly underdeveloped, leaving additional resist residue on the contact area to increase the level of contamination.  The sample is then transferred to a thermal evaporator and pumped down to 10$^{-7}$ torr.  Approximately 35 nm of Cu are then deposited onto the sample, followed by lift-off in acetone.

Each pattern is  designed with many rectangular leads that entirely overlap the sample.  Each lead was 3$\mu$m long in Sample A, and 2$\mu$m in Sample B, while the channel length was varied between 1 and 6$\mu$m in approximately 1$\mu$m increments, similar to the devices found in Ref.~\cite{xia}.  The width of Sample A is $4.7\mu$m and the width of sample B is $.95\mu$m.  The contact resistances were evaluated on a probe station in ambient conditions by measuring the pairs of adjacent contacts on each sample, and then these resistances were plotted vs channel length and normalized by the width of the ribbon, shown in Fig.~\ref{TLM}.  The electric field effect of the channels was tested in vacuum at 4K, and is displayed in Fig.~\ref{gatescans}.  Since the contact resistances derived when the channel is near the Dirac point are very unreliable, as we will explain later, the gate scans are intended to determine the homogeneity of the device properties and the contact resistances are tabulated for the more technologically relevant room temperature case.  

\section{Experimental Results and Discussion}
\begin{figure}[htbp]
\begin{center}
\includegraphics[width=.99\columnwidth]{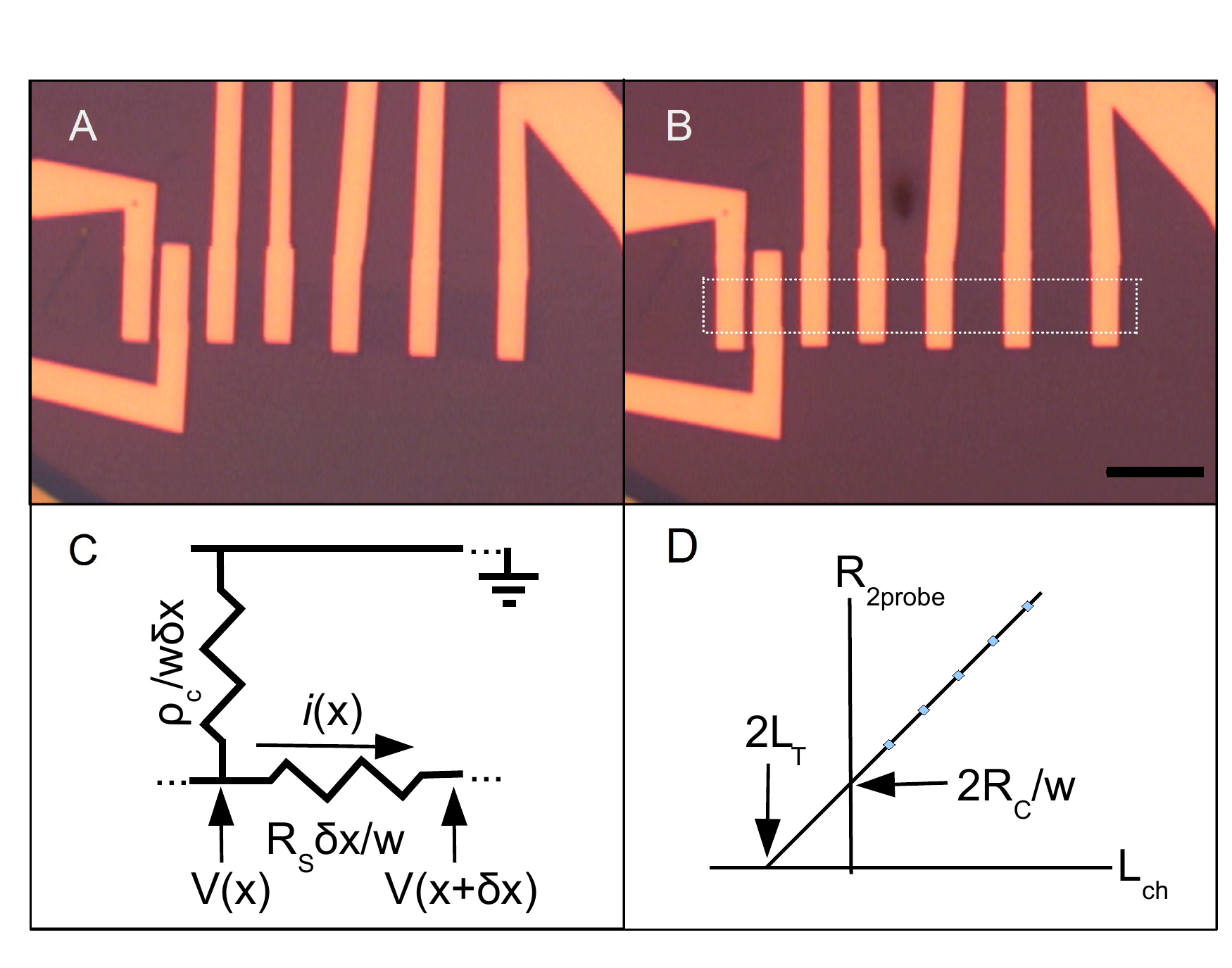}
\caption{Sample A before (A) and after (B) the $306^\circ$C, 12 hour anneal.  There is no damage visible to the device from this treatment.  The scale bar is 10$\mu$m and the device width is 4.7$\mu m$.  (C) A schematic of the relationship of the contact resistivity and the sheet resistance to the distributed resistance network.  (D) A schematic view showing the extraction of the effective contact resistance $R_C$, the sheet resistance $R_S$, and the transfer length $L_T$.}
\label{device}
\end{center}
\end{figure}

Briefly, the TLM analysis describes a contact as a distributed resistance network with a vertical contact resistivity per unit area $\rho_C$, sheet resistance $R_S$, and a perfectly conducting contact metal.  A thorough explanation of the method can be found in Ref.~\cite{bergerhh}.  A set of differential equations for the voltage drop and current flow across the length of such a contact can be derived and solved, allowing one to extract the microscopic parameter $\rho_C$ from the actual measurements.  A schematic of the model is shown in Fig.~\ref{device}-C.

The two probe resistance is given by the equation $R_{2probe}=2R_c/w + (R_S/w)L_{ch}$ where $R_c$ is the effective contact resistance, $R_S$ is the sheet resistance of the graphene channel, $L_{ch}$ is the channel length, and $w$ is the width of the sample.  By plotting the two probe resistance vs channel length, the effective contact resistance (half the y intercept) and sheet resistance (the slope) can be inferred as shown in Fig. 1-D.  A major assumption of the analysis is that the properties of all channels and contacts are uniform.

To test this assumption, Sample A was tested at low temperature to measure two probe resistance vs gate voltage for several channel lengths.  The sample was annealed a third time to reduce environmental dopants on the graphene, and quickly transferred to a dipstick for measurement at 4K.  Fig.~\ref{gatescans} shows that different channels reach the charge neutrality point at similar, but not identical gate voltages.  Lack of consistent properties in large graphene devices is the major source of error in ours and similar studies.

In some respects the TLM is an ideal method to measure graphene contact resistance in the regime of high carrier densities both under the contact and in the channel region, as graphene is atomically thin, there is no averaging over the channel depth, and the sheet resistance under the contact and inside the channel are very similar.  As the sheet resistance in the channel and underneath the contact become significantly different, the method can no longer be used to derive the correct transfer length.~\cite{xia}  Also, resistances due to a potential difference between the contact and graphene region, such as a p-n junction,~\cite{huard} are not accounted for with this method.

The effective contact resistance is related to the resistivity per unit area by $R_c=\sqrt{\rho_cR_S}\coth(d/L_T)$ where d is the length of the contacts, and $L_T=\sqrt{\rho_c/R_S}$.~\cite{bergerhh}  The contacts were fabricated to be much longer than the transfer length, in this case, $\coth(d/L_T)\sim 1$.  The actual measurements support this assumption, deriving transfer lengths less than $.5\mu$m in all samples after vacuum annealing.  Since the sheet resistance used in these calculations is for the contact regions care must be taken when looking at transfer length and contact resistivity, as there is no reason to assume that the sheet resistance in the channel is the same as that in the contact region.

The Cu contacts as deposited have relatively high contact resistances.  To treat this problem we placed them inside a vacuum chamber, pumped down to 10$^{-7}$ torr and annealed them for 12-15 hours.  Sample A was annealed twice, once at 260$^\circ$C, and again at 306$^\circ$C.  The second sample was annealed only once at 300$^\circ$C.  After annealing, the contact resistance in general decreases, even when the sample is heavily contaminated as in Sample B.  Shorter times were tried with varying results.  The extremely long bake time consistently improves the contact resistance.  A method involving gentle O$_2$ plasma exposure in combination with a shorter, higher temperature anneal was recently explored.~\cite{robinson}  We find that our derived values for the contact resistivity on the order of $100\Omega\mu m^{2}$ compares favorably with their method for Cu contacts with oxygen plasma treatment $1000 \Omega\mu m^{2}$.  Note that O$_2$ plasma is not used to clean the graphene surface from contamination in the technique presented here.  We also find that our contact resistance also compares favorably to Pd/Pt contacts reported in Ref.~\cite{xia}, reporting effective contact resistances at high carrier densities of $110\Omega\mu m$.

\begin{figure}[htbp]
\begin{center}
\includegraphics[width=.99\columnwidth]{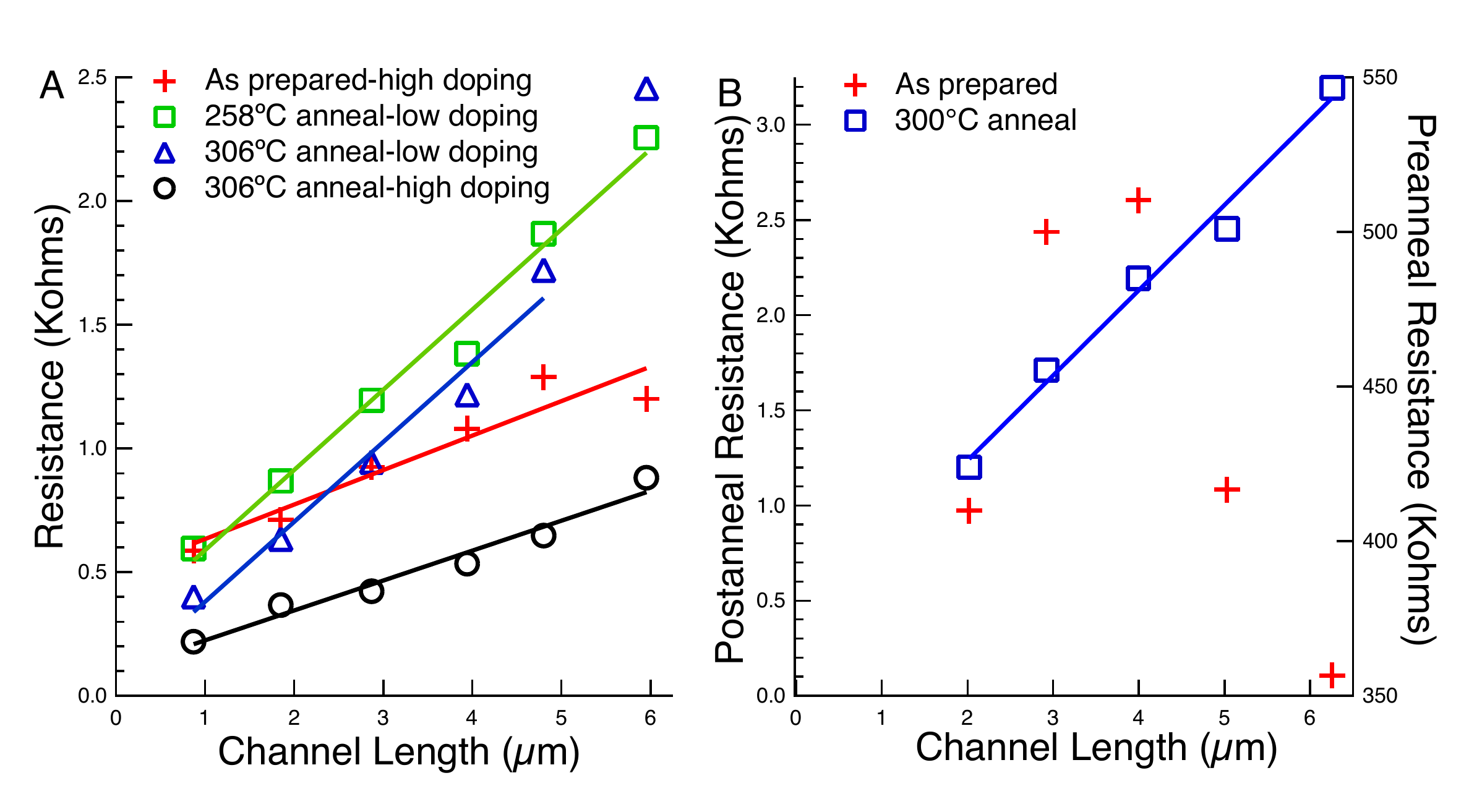}
\caption{A: Sample A after several annealing cycles to the sample. Immediately after removal from the vacuum chamber, the TLM measurement displays a much larger uncertainty than after the channel has returned to a high doping value, after being exposed in air for $>$100 hours. The point for the longest channel length for the 306$^{\circ}$C anneal at low doping is omitted from the fit, as it causes the effective contact resistance to become negative. B: Sample B, due to the additional contamination, the resistance is dominated by the contacts, and is unrelated to the channel length, however; the contact resistance decreases dramatically after vacuum annealing.}
\label{TLM}
\end{center}
\end{figure}

\begin{figure}[htbp]
\begin{center}
\includegraphics[width=.99\columnwidth]{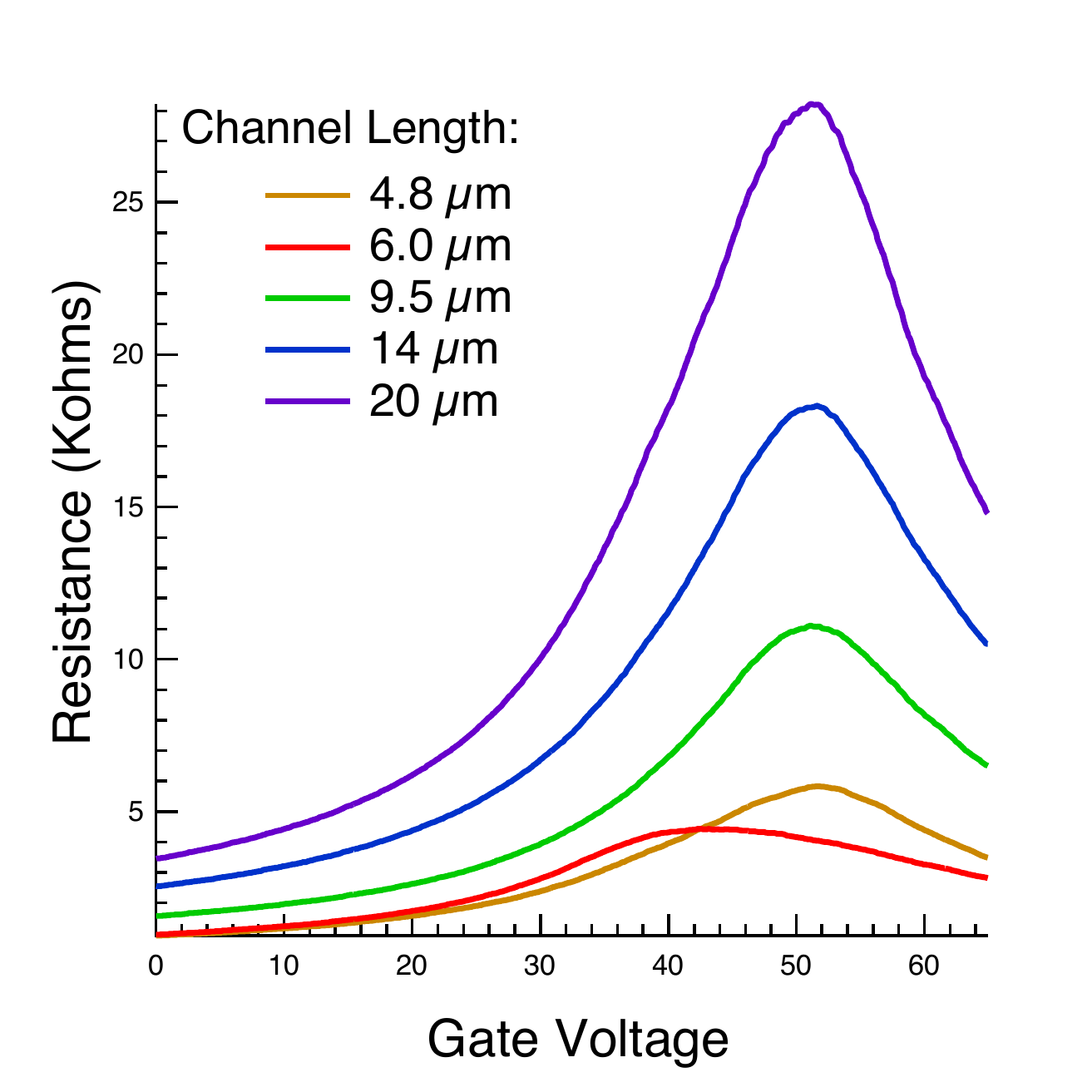}
\caption{Gate voltage dependence at 4K for several graphene transistors on Sample A.  The properties can be seen to be largely, but not perfectly uniform, adding uncertainties to the contact resistance measurements.  Note that for some channel lengths shown the current passes beneath contacts that are allowed to float.}
\label{gatescans}
\end{center}
\end{figure}

Fig.~\ref{TLM}-A displays the line fits to the two probe resistances for Sample A.  A clear improvement can be seen after the annealing process.  Fig~\ref{TLM}-B displays the line fit to the two probe resistances for Sample B.  Due to the higher contamination of this sample the two probe resistances were hundreds of k$\Omega$s, however; the annealing process still repaired the leads.  The extracted values of $R_c$, $L_T$, $R_S$, and $\rho_c$ for both samples are displayed in Table I.  

A sample not shown here further demonstrates the effectiveness of annealing to repair imperfect graphene contacts.  It was dipped in dilute (12\%) HNO$_3$ for several seconds for a different experiment.  The leads were not fully oxidized by this process, and in fact most still made contact to the graphene via k$\Omega$ resistances.  After several weeks in atmosphere the contacts were all highly resistive ($>$M$\Omega$s), however; an anneal such as the one described above returned the contacts to resistances on the order of k$\Omega$, and some tarnishing visible before the anneal was no longer visible afterwards.  We do not have precise contact resistance values for this sample since its geometry was not designed for the types of measurements presented here.

It is interesting that the method presented here shows good quality contacts with a metal theoretically predicted to have poor bonding with graphene.  The TLM analysis clearly shows that the effective contact resistance $R_C$ actually measured is due to both $\rho_C$ and $R_S$.  The sp$_2$ bonding in the graphene lattice allows the unhybridized p$_z$ orbitals to form a conduction band, making graphene a semimetal.  This should be contrasted with the sp$_3$ bonding in diamond which leads to a wide band gap semiconductor.  Therefore, though they may decrease $\rho_C$, a bonding contact can lead to a lower sheet resistance in the graphene underneath due to the increased sp$_3$ bonding.  Given these arguments, it is not immediately obvious what metal will create the lowest effective contact resistance, and non-bonding metals may merit more investigation for practical applications as part of a systematic search for the ideal metal contact.

In conclusion we present a method to significantly improve the contact resistance between graphene and Cu.  Though the technique presented here requires long annealing times and high vacuum, there is no reason to believe that the use of flash annealing, or other high temperature methods may not significantly decrease the time necessary to anneal the contacts.  We also find that the annealing process has the ability to improve contact resistance even in the presence of large degrees of contamination.

Funding from DOE grant DE-FG02-06ER26381

\newpage

\begin{table*}
\label{Calculated}
\begin{center}
\begin{tabular}{|l|c|c|c|c|}
\hline
Sample & $R_c (\Omega\mu m)$ &$L_T (\mu m)$&$R_S (\Omega/\square)$&$\rho_c (\Omega\mu m^2)$\\
\hline
A - as prepared &1161& 1.77&655 &2058\\
\hline
A - 260$^\circ$C anneal &620&0.407&1524&252\\
\hline
A - 306$^\circ$C anneal &&&&\\
high doping &241&0.424&568&102\\
low doping &135&0.089&1516&12.0\\
\hline
B - as prepared &1.75e6&N/A&N/A&5.26e6\\
\hline
B - 300$^\circ$C anneal &163&0.384&425&213\\
\hline
\end{tabular}
\caption{Measured values of $R_c$ and $R_S$, along with derived values of $L_T=R_C/R_S$ and $\rho_c=R_C^2/R_S$}
\end{center}
\end{table*}

\end{document}